\documentclass[12pt,thmsa]{article}
\usepackage{sw20lart}

\input{tcilatex}
\begin{document}

\begin{center}
{\LARGE A polynomial-time solution to the parity problem on an NMR quantum
computer}

\ 

Xijia Miao

Correspondence address: Center for Magnetic Resonance Research, University
of Minnesota, Minneapolis, MN 55455, USA; E-mail: miao@cmrr.umn.edu

\ 

Abstract
\end{center}

An efficient quantum algorithm is proposed to solve in polynomial time the
parity problem, one of the hardest problems both in conventional quantum
computation and in classical computation, on NMR quantum computers. It is
based on the quantum parallelism principle in a quantum ensemble, the
selective decoherence manipulation, and the NMR phase-sensitive measurement.
The quantum circuit for the quantum algorithm is designed explicitly. 
\textbf{\newline
}\newline
\textbf{1. Introduction}

It is well known that the parity problem is one of the hardest problems [1,
2]. These hard problems [1-4] including the parity problem, majority
problem, and some iteration problem, etc., can not be solved efficiently
both on the classical and quantum computers in polynomial time so far. It
has been shown that quantum computers can achieve no more than a polynomial
speedup over the classical computers in solving these hard problems,
implying that a quantum computer cannot outperform a classical computer in
solving these problems [1, 2]. On the other hand, since nuclear magnetic
resonance (NMR) technique [5, 6] has been proposed to implement in
experiment quantum computation [7, 8], there are some difficulties, limits,
and questions in NMR quantum computation [9, 10], although a great and rapid
progress has been achieved and a great flow of the work has been reported in
past several years. Due to too low spin polarization in a spin ensemble at
room temperature [9] it can not be certain up to now from quantum mechanical
principle whether NMR quantum computation is a real quantum computation and
can be more powerful than the classical computation. Some peoples [11-13]
even claimed from the view point of the quantum entanglement theory that NMR
quantum computers are not in fact real quantum computers, and suspect
whether or not NMR quantum computation based on the effective pure states
[7, 8] is as powerful as the quantum computation based on the pure quantum
states also owing to the two main drawbacks, that is, an exponent reduction
of NMR signal intensity as the qubit number and a high cost for the
preparation of the initial effective pure state on NMR quantum computation.
However, most recently an efficient quantum algorithm running on the NMR
quantum computers is proposed to solve in polynomial time the unsorted
quantum search problem [14]. It is well known that the conventional quantum
search algorithm, i.e., the Grover algorithm, based on the pure quantum
states [15] or the effective pure states [16] cannot solve the search
problem in polynomial time, although it is a quadric speedup algorithm over
the classical search algorithms. Therefore, the new quantum search algorithm
shows that quantum algorithms on the NMR quantum computers are not only as
powerful as on the conventional quantum computers based on the pure quantum
states but also some quantum computations running on the NMR quantum
computer can be more powerful than running on the conventional one. This
algorithm is based on the quantum parallelism principle in quantum
ensembles, the selective decoherence manipulation, and NMR phase sensitive
measurement [14]. In this paper I further show that there is an efficient
quantum algorithm to solve the parity problem in polynomial time on the NMR
quantum computers. It is still based on the quantum parallelism principle in
quantum ensembles, the selective decoherence manipulation, and NMR phase
sensitive measurement. The present quantum algorithm to efficiently solve
the parity problem in polynomial time supports powerfully the belief that
the NP hard problems can be efficiently solved in polynomial time on quantum
computers and quantum computers can outperform eventually the classical
computers. \newline
\newline
\textbf{2. Quantum parallelism in a quantum ensemble}

The quantum parallelism is one of the most basic principles in quantum
computation [17]. In principle, the quantum parallelism offers the
possibility that quantum computation can be more powerful than the classical
computation, although it is usually difficult to extract the desired
computational output from a quantum system in quantum computation directly
by the quantum mechanical measurement [18]. The quantum parallelism may be
understood simply through the process that superposition of a quantum system
is acted on by a unitary transformation of quantum computation. Since a
quantum system can be at any superposition which can be a linear sum of the
usual quantum computational bases, while number and functional operation in
mathematics usually may be represented by the computational bases and the
unitary transformation, respectively, in quantum computation, then
performing quantum mechanically the unitary transformation of the functional
operation on the superposition means really that the functional operation is
performed in a parallel form with all input numbers represented by those
computational bases of the superposition. Because quantum computation is a
reversible computational process [19] an arbitrary quantum state of the
quantum system is equivalent unitarily to a computational basis or any other
quantum state of the system, that is, any pair of quantum states of the
system can be unitarily converted completely each other in quantum
computation [20] when there are not quantum measurement and decoherence
process during the quantum computation. Now, if the superposition is
expressed as a unitary transformation acting on some usual computational
basis, the above functional operation process really can start at the
computational basis instead of the superposition. However, this does not
mean that there is not the quantum parallelism in the computational process
even when the initial input state is not superposition. This is because
superposition of the quantum system is usually created during the unitary
transformation of the functional operation. Therefore, in fact the quantum
parallelism is also characterized by unitary transformation besides
superposition. Due to this fact sometime the initial input state is included
in a quantum algorithm [21]. However, it is key important to separate the
initial input state representing in mathematics number and the unitary
transformation representing the functional operation mathematically when the
quantum parallelism principle is extended to a quantum ensemble and can be
able to play an important role in quantum computation. In general, a
molecule spin system and its spin ensemble composed of the molecule spin
system obey the unitary dynamics governed by the same spin Hamiltonian in
nuclear magnetic resonance spectroscopy [5, 6]. This point is very helpful
for quantum computation making a transition from the quantum spin system
version to the spin ensemble version. It has been shown that any unitary
transformation of quantum computation can be constructed with the spin
Hamiltonian of a spin system [22]. Then the unitary transformation is still
available and takes the same form when quantum computation is carried out in
the spin ensemble with the same spin Hamiltonian of the molecule spin
system. The significant difference between the spin system and its ensemble
is that there are mixed states in addition to the pure quantum states in the
spin ensemble. However, just due to the fact that a spin ensemble can be in
a mixed state, the spin ensemble can provide the quantum parallelism
principle with a much larger space for it to bring into play in quantum
computation when the unitary transformation and the initial input state are
independent on each other in quantum computation. This is because the
initial input state of the quantum computation may be any mixed state in
addition to the pure quantum state in a spin ensemble.

In a quantum system with $n$ qubits there are $N=2^{n}$ usual computational
bases or $N$ any other eigenbases in the $N-$dimensional Hilbert space of
the quantum system, and any superposition of the quantum system can be
expressed as a linear sum of the $N$ eigenvectors in the vector
representation. In the matrix representation any superposition of the
quantum system is still expressed as a linear sum of $N$ linearly
independent quantum states which are now represented by matrices, but this
matrix representation just gives the same information of the quantum system
as the vector representation provides. For example, the effective pure state
in a spin ensemble [7, 8] provides a faithful matrix representation for the
pure quantum state of the spin system in the spin ensemble, but it does not
give more information than the pure quantum state in quantum computation.
However, any superposition of a spin ensemble, that is, any mixed state or
any density operator of a spin ensemble must be expanded as a sum of any $%
4^{n}$ linearly independent matrix bases in the $4^{n}-$dimensional
Liouville space of the spin ensemble [5]. A mixed state is not always able
to be unitarily transferred completely into another mixed state in a spin
ensemble, and this is quite different from the case in a quantum system in
quantum computation. If any unitary transformation and the initial input
state are independent on each other in a quantum algorithm whose quantum
circuit is constructed with the spin Hamiltonian of a spin system or its
spin ensemble, any mixed state of the spin ensemble can be used as the
initial input state of the quantum algorithm run on the spin ensemble. Then,
the initial input state of the quantum algorithm can be selected from the
larger $4^{n}-$dimensional Liouville space of the spin ensemble instead of
the smaller $2^{n}-$dimensional Hilbert space of the spin system. This could
be helpful to design efficient quantum algorithms to solve some hard
problems on a spin ensemble.

How to solve mathematical problems in a spin ensemble like in a quantum
system? The well-known scheme is based on the effective pure state [7, 8],
but according to the scheme the mixed states in a spin ensemble really do
not play an important role in the quantum computation. A possible scheme
proposed here is that the complete information of the problems to be solved
first is contained or hidden in unitary transformation of quantum
computation or its corresponding effective Hamiltonian and / or the mixed
states in the spin ensemble, and the simpler and better one is that only the
unitary transformation contains the complete information of the problem to
be solved. By choosing the suitable initial mixed states one runs the
quantum algorithm on the spin ensemble, and then extracts the desired
information of the problem by NMR phase-sensitive measurement technique [5,
6]. The latter scheme has been used to find the efficient quantum search
algorithm on NMR quantum computers [14]. \newline
\newline
\textbf{3. Selective decoherence manipulation }

All the coherent and the noncoherent components (the longitudinal
magnetization and spin order components in nonequilibrium state) in a spin
ensemble decay irreversibly as time and at the same time the spin ensemble
returns to the thermal equilibrium state irreversibly, but their decay rates
are quite different [5, 6]. For example, in a spin ensemble the longitudinal
relaxation time is always longer than the transverse relaxation time, and
for some macromolecules the longitudinal relaxation time can be even hundred
and thousand times longer than the transverse relaxation time. The important
fact is that the decay rates for some coherent and noncoherent components in
a spin ensemble can be manipulated and controlled externally. For example,
the nonzero-order multiple-quantum coherences including the transverse
magnetizations are much sensitive to the inhomogeneous external magnetic
field but the longitudinal magnetizations are not [5, 6]. Then one can make
use of the inhomogeneous external magnetic field, for example, a gradient
magnetic field [6], to control the decay of these multiple-quantum
coherences, so that at a short time these coherent components are cancelled
but only the zero-quantum coherences and the longitudinal magnetizations and
spin order components almost keep unchanged. Finally, the residual
zero-quantum coherences can be cancelled by the zero-quantum dephasing pulse
[24] or z-filter [5, 6, 25] so that only the longitudinal magnetization and
spin order components are retained selectively in the spin ensemble. Such
selective decoherence manipulation has been used extensively in nuclear
magnetic resonance experiments. In quantum computation the quantum
measurement is also a selective decoherence manipulation. The quantum state
measured is retained but any other states decay to zero quickly in a quantum
system when performing quantum measurement. The importance is how to exploit
in a positive manner the selective decoherence manipulation to help to find
new polynomial-time quantum algorithms in quantum computation. In quantum
computation the final goal is to extract the desired information of the
problem to be solved by a quantum algorithm from the quantum system, but one
need not obtain the complete information of the quantum system. Therefore,
one may use properly the selective decoherence manipulation to cancel the
undesired information but obtain valuable information which is sufficient to
solve the problem. The spin ensemble provides a very suitable place to
perform the selective decoherence manipulation for quantum computation since
nuclear magnetic resonance technique is well-developed at present. The
massive quantum parallelism makes it possible for quantum computation
outperforming classical one, but the desired computational output could not
be obtained easily due to the limit of the quantum measurement principle in
a quantum system. However, since the selective decoherence manipulation in a
spin ensemble can be performed more conveniently, the undesired coherent and
noncoherent components created in quantum computation in a spin ensemble
could be cancelled and as a consequence, the desired computational results
could be detected conveniently by the NMR phase-sensitive measurement
technique [14]. Here, the selective decoherence manipulation is a positive
effect in quantum computation. Though the decoherence effect is usually a
negative effect on quantum computation and need to be overcome, the
selective decoherence manipulation could be useful to help finding new
efficient quantum algorithms. For example, most recently an efficient
quantum search algorithm [14] based on the selective decoherence
manipulation is proposed on the NMR quantum computers. \newline
\newline
\textbf{4. The quantum algorithm to solve efficiently the parity problem}%
\newline

The parity problem can be outlined briefly below. Given a function $f(x)$
which is defined on the integers $x$: $0\leq x\leq N-1$ $(N=2^{n}).$ For any 
$x$ $(0\leq x\leq N-1)$ the function $f(x)$ can only take either $+1$ or $-1$%
. The parity of the function $f(x)$ can be defined as [1]

$\qquad \qquad \qquad \qquad \qquad par(f)=\stackrel{N-1}{\stackunder{x=0}{%
\prod }}f(x)\qquad \qquad \qquad \qquad \qquad \qquad \ \ (1)$ \newline
where $f(x)=\pm 1,\ x=0,1,...,N-1.$ The parity problem of the function $f(x)$
is how to determine the parity function $par(f)$. Obviously, the parity
function $par(f)$ can only take either +1 or -1. To determine the parity
problem in classical computation one needs to call $N$ times the function $%
f(x)$ so that $f(x)$ can be known for each $x$ value and hence the parity
function $par(f)$ can be determined from Eq.(1). In quantum computation
evaluation of the function $f(x)$ over all the $x$ values $0\leq x\leq N-1$
can be performed simultaneously by only one call of the functional operation
of $f(x)$ on the superposition of the quantum system due to the quantum
parallelism, but the final result, that is, all values of the function $f(x)$%
, can not be extracted simultaneously due to the limit of quantum mechanical
measurement. As a result, the cost to solve the parity problem on a quantum
computer is not really much lower than on a classical computer, implying
that quantum computers can not outperform classical computers in determining
the problem. However, the parity problem is a global-type or collective-type
problem but not an individual one of the function $f(x).$ Perhaps people
could not evaluate each value of the function $f(x)$ over all $x$ values but
can determine the parity function (1).

First of all, the unitary transformation corresponding evaluation of the
function $f(x)$ is introduced. This unitary operation is the oracle unitary
operation in the quantum algorithm to solve the parity problem. Assume that
the evaluation of the function $f(x)$ can be expressed by performing the
oracle unitary operation $U_{f}$ on any quantum state $|x\rangle |S\rangle $
in a quantum system$,$

$\qquad \qquad U_{f}|x\rangle |S\rangle =f(x)|x\rangle |S\rangle =\exp
(-i\pi g(x))|x\rangle |S\rangle \qquad \qquad \qquad \quad (2)$ \newline
where the auxiliary quantum state $|S\rangle $ must be chosen properly [1,
14, 18] so that equation (2) holds, and the function $g(x)$ is defined by $%
f(x)=\exp (-i\pi g(x)),$ that is, $g(x)=0$ if $f(x)=1$; $g(x)=1$ if $f(x)=-1$%
. When the oracle unitary operation $U_{f}$ acts once on any superposition
of the quantum system all the values of the function $f(x)$ over all $x$: $%
0\leq x\leq N-1$ are evaluated simultaneously,

$\qquad \qquad \qquad U_{f}\stackrel{N-1}{\stackunder{x=0}{\sum }}%
a_{x}|x\rangle |S\rangle =\stackrel{N-1}{\stackunder{x=0}{\sum }}%
a_{x}f(x)|x\rangle |S\rangle .\qquad \qquad \qquad \qquad \quad (3)$ \newline
However, these function values are not simultaneously observable in the
quantum system due to the limit of the quantum mechanical measurement. It
has been shown that to determine certainly the parity problem on a quantum
computer one needs to execute around $N/2$ times the oracle unitary
operation $U_{f}$ [1, 2]. Equation (3) shows that the oracle unitary
operation $U_{f}$ is independent upon any quantum state $|x\rangle $ of the
work qubits $I$ (hereafter the symbol $I$ denotes the work qubits) in the
quantum system but related to the auxiliary quantum state $|S\rangle $
(hereafter the symbol $S$ denotes the auxiliary qubits)$.$

In order to construct the explicit equivalent form for the oracle unitary
operation $U_{f}$ one may use the selective phase-shift operations $%
C_{s}(\theta )$ to express the oracle unitary operation $U_{f}$. The
selective phase-shift operation $C_{s}(\theta )$ is a diagonal unitary
operator of the longitudinal magnetization and spin order operator subspace
and defined as [14, 22]

$\qquad \qquad \qquad \qquad C_{s}(\theta )=\exp (-i\theta D_{s})\qquad
\qquad \qquad \qquad \qquad \quad \ \ \qquad (4)$ \newline
where the diagonal operator $D_{s}$ is defined as $%
D_{s}=diag(0,...,0,1,0,...,0),$ that is, $(D_{s})_{ss}=1$ for index $s$ and $%
(D_{s})_{rr}=0$ for any other index $r\neq s.$ Note that when the selective
phase-shift operation $C_{x}(\theta )$ acts on any quantum state $|r\rangle
|S\rangle $ a phase shift exp$(-i\theta )$ is created only when the index $%
r=x$, otherwise the quantum state $|r\rangle |S\rangle $ keeps unchanged,

$C_{x}(\pi g(x))|r\rangle |S\rangle =\exp (-i\pi g(x)\delta _{rx})|r\rangle
|S\rangle =\{ 
\begin{array}{l}
|r\rangle |S\rangle \qquad \quad \text{if }r\neq x \\ 
f(x)|x\rangle |S\rangle \quad \text{if }r=x
\end{array}
,\qquad $\newline
and also note that any pair of the selective phase-shift operations commute
each other. It can be seen from Eqs.(2) and (3) that the oracle unitary
operation $U_{f}$ is really equivalent to the product of all $N$ selective
phase-shift operations $C_{x}(\pi g(x)),$ $x=0,1,...,N-1,$

$\qquad \qquad \qquad \qquad U_{f}=\stackrel{N-1}{\stackunder{x=0}{\prod }}%
C_{x}(\pi g(x)).\qquad \qquad \qquad \qquad \qquad \qquad \quad \ \ (5)$ 
\newline
In a more general case, in the present quantum algorithm the oracle unitary
operation $U_{f}$ of Eq.(5) can be replaced with the more general oracle
unitary operation $U_{o}(\theta )$ which is the product of the $N$ selective
phase-shift operations $C_{x}(\theta g(x)),$ $x=0,1,...,N-1,$ with any phase
angle $\theta $ by choosing properly the auxiliary quantum state $|S\rangle
=|0\rangle |1\rangle $ [14]

$\qquad \qquad U_{o}(\theta )=U_{f}V(\theta )U_{f}=$ $\stackrel{N-1}{%
\stackunder{x=0}{\prod }}C_{x}(\theta g(x)).\qquad \qquad \qquad \qquad
\qquad \ \ (6)$ \newline
Note that the function $g(x)=0,1$ for $f(x)=1,-1$, respectively, as shown in
Eq.(2). The parity function $par(f)$ of Eq.(1) can be expressed as

$par(f)=\stackrel{N-1}{\stackunder{x=0}{\prod }}f(x)=\stackrel{N-1}{%
\stackunder{x=0}{\prod }}\exp (-i\pi g(x))=\exp (-i\pi \stackrel{N-1}{%
\stackunder{x=0}{\sum }}g(x)).$ \qquad $\qquad (7)\newline
$Define the integer phase parameter $G$ as

$\qquad \qquad \qquad \qquad G=\stackrel{N-1}{\stackunder{x=0}{\sum }}g(x).$
\qquad $\qquad \qquad \qquad \qquad \qquad \qquad \qquad \ (8)$\newline
Obviously, $par(f)=+1$ if the integer phase parameter $G$ is an even number
and $par(f)=-1$ if $G$ is an odd number. Therefore, the parity problem is
really determined completely by the parity of the integer phase parameter $G$%
. Obviously, the integer phase parameter $G$ is a global or collective
quantity. Its parity may be determined without knowing each individual value
of the function $f(x)$ and hence the parity problem of the function $f(x)$
could be solved efficiently without knowing all the functional values $f(x)$
in advance.

In the above the oracle unitary operations $U_{f}$ of the functional
operation $f(x)$ and $U_{o}(\theta )$ are introduced in a quantum system.
Now to extend these oracle unitary operations to the ensemble composed of
the quantum system the auxiliary quantum states $|S\rangle $ in the quantum
system must be kept unchanged or be replaced by their corresponding
equivalent effective pure states [23] in the quantum ensemble because the
oracle unitary operations are related to the auxiliary quantum states $%
|S\rangle ,$ as can be seen in Eqs.(2), (3) and (6), but the quantum state $%
|x\rangle $ of the work qubits $I$ can be replaced by any mixed state or
density operator of the ensemble [14].

To describe conveniently the evolution process of the quantum spin system or
its spin ensemble under the action of the oracle unitary operation $%
U_{o}(\theta )$ (6) one needs to exploit the following general unitary
transformation when the selective phase-shift operation $C_{s}(\theta )$
acts on any density operator $\rho _{I}(0)$ of the work qubits $I$ of the
spin ensemble [26]:

$\qquad C_{s}(\theta )\rho _{I}(0)C_{s}(\theta )^{-1}=\rho _{I}(0)-(1-\cos
\theta )[\rho _{I}(0),D_{s}]_{+}$

$\qquad \qquad \qquad +i\sin \theta [\rho _{I}(0),D_{s}]+[(1-\cos \theta
)^{2}+\sin ^{2}\theta ]D_{s}\rho _{I}(0)D_{s}$ \qquad $(9)$ \newline
where the commutation $[\rho _{I}(0),D_{s}]_{+}=\rho _{I}(0)D_{s}+D_{s}\rho
_{I}(0)$ and $[\rho _{I}(0),D_{s}]=\rho _{I}(0)D_{s}-D_{s}\rho _{I}(0).$
With the help of the unitary transformation (9) and the explicit expression
(6) of the oracle unitary operation $U_{o}(\theta )$ one can obtain easily
the following unitary transformation when the oracle unitary operation $%
U_{o}(\theta )$ acts on any density operator $\rho _{I}(0)|S\rangle \langle
S|$ of the ensemble

$U_{o}(\theta )\rho _{I}(0)U_{o}(\theta )^{-1}=\rho _{I}(0)$

$-(1-\cos \theta )[\rho _{I}(0),\stackrel{N-1}{\stackunder{s=0}{\sum }}%
g(s)D_{s}]_{+}+i\sin \theta [\rho _{I}(0),\stackrel{N-1}{\stackunder{s=0}{%
\sum }}g(s)D_{s}]$

$+[(1-\cos \theta )^{2}+\sin ^{2}\theta ]\stackrel{N-1}{\stackunder{s=0}{%
\sum }}\stackrel{N-1}{\stackunder{t=0}{\sum }}g(s)g(t)D_{s}\rho _{I}(0)D_{t}$%
.\qquad $\qquad \qquad \qquad \quad \ (10)$ \newline
Here, without confusion the auxiliary pure quantum state or effective pure
state $|S\rangle \langle S|$ is omitted. Equation (10) is the basic unitary
transformation to analyze the quantum algorithm to solve efficiently the
party problem on an NMR quantum computer. The initial density operator $\rho
_{I}(0)$ must be related to each quantum state $|x\rangle $ $(x=0,1,..,$ $%
N-1)$, otherwise it keeps unchanged under the action of the selective
phase-shift operation $C_{x}(\theta )$, $C_{x}(\theta )\rho
_{I}(0)C_{x}(\theta )^{-1}=\rho _{I}(0),$ and as a result, one may not
determine certainly the parity function of Eq.(1). One of the proper initial
density operators in a spin ensemble may be

$\qquad \qquad \qquad \qquad \rho _{I}(0)=\stackrel{n}{\stackunder{k=1}{\sum 
}}\varepsilon _{k}I_{ky}$ \qquad \qquad \qquad \qquad \qquad \qquad \qquad $%
\ \ (11)$\newline
where $\varepsilon _{k}$ is the spin polarization parameter of spin $k$ and
the term proportional to the unity operator is omitted without losing
generality. It has proved that the density operator $\rho _{I}(0)$ of
Eq.(11) are related to all the quantum states $|x\rangle $ of the spin
system in the spin ensemble [14]. Furthermore, it is convenient to employ
the unit-number vector representation $\{a_{k}^{s}\}$ [14] to analyze the
unitary transformation (10). In the unit-number vector representation the
diagonal operator $D_{s}$ can be expressed as [14]

$\qquad D_{s}=(\frac{1}{2}E_{1}+a_{1}^{s}I_{1z})\bigotimes ...\bigotimes (%
\frac{1}{2}E_{k}+a_{k}^{s}I_{kz})\bigotimes ...\bigotimes (\frac{1}{2}%
E_{n}+a_{n}^{s}I_{nz}).$\qquad $(12)$\newline
Assume that the work qubits $I$ are prepared initially at the density
operator of Eq.(11) and the auxiliary qubit $S$ is in the quantum state $%
|S\rangle \langle S|$ [23] in the spin ensemble. Then make the oracle
unitary operation of $U_{o}(\theta )$ on the spin ensemble. The spin
ensemble will evolve according to the following unitary transformation
obtained easily from Eq.(10):

$U_{o}(\theta )\stackrel{n}{\stackunder{k=1}{\sum }}\varepsilon
_{k}I_{ky}U_{o}(\theta )^{-1}=\stackrel{n}{\stackunder{k=1}{\sum }}%
\varepsilon _{k}I_{ky}$

$-$ $(1-\cos \theta )\stackrel{N-1}{\stackunder{s=0}{\sum }}g(s)\stackrel{n}{%
\stackunder{k=1}{\sum }}(\frac{1}{2}E_{1}+a_{1}^{s}I_{1z})\bigotimes ...$

$\bigotimes (\frac{1}{2}E_{k-1}+a_{k-1}^{s}I_{k-1z})\bigotimes (\varepsilon
_{k}I_{ky})\bigotimes (\frac{1}{2}E_{k+1}+a_{k+1}^{s}I_{k+1z})\bigotimes
...\bigotimes (\frac{1}{2}E_{n}+a_{n}^{s}I_{nz})$

$-\sin \theta \stackrel{N-1}{\stackunder{s=0}{\sum }}g(s)\stackrel{n}{%
\stackunder{k=1}{\sum }}(\frac{1}{2}E_{1}+a_{1}^{s}I_{1z})\bigotimes ...$

$\bigotimes (\frac{1}{2}E_{k-1}+a_{k-1}^{s}I_{k-1z})\bigotimes (\varepsilon
_{k}a_{k}^{s}I_{kx})\bigotimes (\frac{1}{2}E_{k+1}+a_{k+1}^{s}I_{k+1z})%
\bigotimes ...\bigotimes (\frac{1}{2}E_{n}+a_{n}^{s}I_{nz})$

$+[(1-\cos \theta )^{2}+\sin ^{2}\theta ]\stackrel{N-1}{\stackunder{t>s=0}{%
\sum }}g(s)g(t)\stackrel{n}{\stackunder{k=1}{\sum }}[\frac{1}{4}%
(1+a_{1}^{s}a_{1}^{t})E_{1}+\frac{1}{2}(a_{1}^{s}+a_{1}^{t})I_{1z}]%
\bigotimes ...$

$\bigotimes [\frac{1}{4}(1+a_{k-1}^{s}a_{k-1}^{t})E_{k-1}+\frac{1}{2}%
(a_{k-1}^{s}+a_{k-1}^{t})I_{k-1z}]\bigotimes [\frac{1}{2}\varepsilon
_{k}(1-a_{k}^{s}a_{k}^{t})I_{ky}]$

$\bigotimes [\frac{1}{4}(1+a_{k+1}^{s}a_{k+1}^{t})E_{k+1}+\frac{1}{2}%
(a_{k+1}^{s}+a_{k+1}^{t})I_{k+1z}]\bigotimes ...$

$\bigotimes [\frac{1}{4}(1+a_{n}^{s}a_{n}^{t})E_{n}+\frac{1}{2}%
(a_{n}^{s}+a_{n}^{t})I_{nz}].\qquad \qquad \qquad \qquad \qquad \qquad
\qquad \ \ (13)$ \newline
The mixed state of Eq.(13) of the spin ensemble after the action of the
oracle unitary operation of $U_{o}(\theta )$ is quite complicated, but it
can be simplified by the selective decoherence manipulation. First, a hard
90 degree pulse along y-axis direction is applied to all the spins of the
work qubits $I$ of the spin ensemble and then a purge pulse [5, 6, 24, 25],
that consists of the z-gradient field used to cancel all the nonzero
multiple-quantum coherences including single-quantum coherences, and the
zero-quantum dephasing pulse [24] used to cancel all the zero-quantum
coherences, is applied to cancel all the undesired multiple-quantum
coherences but only the longitudinal magnetization and spin order components
keep unchanged. Then the final density operator is reduced to the form

$\quad \rho _{f}=(2/N)\stackrel{n}{\stackunder{k=1}{\sum }}\varepsilon
_{k}I_{kz}(\stackrel{N-1}{\stackunder{s=0}{\sum }}g(s)a_{k}^{s})=(2/N)%
\stackrel{n}{\stackunder{k=1}{\sum }}\varepsilon _{k}P_{k}I_{kz}\qquad
\qquad \quad \ \ (14)$ \newline
where the phase angle $\theta $ in the oracle unitary operation $%
U_{o}(\theta )$ is taken as 90 degree. The density operator $\rho _{f}$ of
Eq.(14) determines the final observable NMR signal of any spin $k$ $%
(k=1,2,...,n)$ of the spin system in the spin ensemble, and the intensity of
the NMR signal of the spin $k$ is proportional to the sum $P_{k},$ which is
an integer defined by

$\qquad \qquad \qquad P_{k}=(\stackrel{N-1}{\stackunder{s=0}{\sum }}%
g(s)a_{k}^{s})=\stackrel{G}{\stackunder{t}{\sum }}a_{k}^{t}$ \qquad $\qquad
\qquad \qquad \ \ \qquad \qquad (15)$\newline
where the latter equality holds due to the fact that the function $g(s)=0,1$
for any index $s$, and the second sum (the index $t$) runs over only those
indexes $s$ with $g(s)=1$. It turns out easily that the integer $P_{k}$ can
take any integer from $-N/2$ to $N/2$ since $a_{k}^{t}=\pm 1$ for any
indexes $k$ and $t,$ and a half of all $N$ unity numbers $\{a_{k}^{l}\}$
take $+1$ for any given index $k.$ It follows from Eq.(15) that the parity
of the integers $P_{k}$ $(k=1,2,...,n)$ is determined uniquely by that of
the integer phase parameter $G$ and in turn, the parity of the integer phase
parameter $G$ is also determined uniquely by that of the integer $P_{k}$,
that is, if the integer phase parameter $G$ is an even number each integer $%
P_{k}$ must also be an even number, and if $G$ is an odd number each $P_{k}$
is an odd number too. Now, if every integer $P_{k}$ is an odd number it
follows from the density operator $\rho _{f}$ of Eq.(14) that the NMR signal
for any spin $k$ $(k=1,2,...,n)$ of the spin system in the ensemble always
has a nonzero intensity. Obviously, if there is zero-intensity NMR signal of
the density operator $\rho _{f}$ of Eq.(14) for any arbitrary spin $k$ of
the spin system in the ensemble it can be certain that all the integers $%
\{P_{k}\}$ are even numbers (including zero) and hence the integer phase
parameter $G$ is also an even number. However, there is another possibility
that the NMR signal of the density operator $\rho _{f}$ of Eq.(14) for all
the spins in the ensemble can have nonzero intensity even when all the
integers $\{P_{k}\}$ are nonzero even numbers. It is clear that the parity
problem can be solved efficiently by the above process only when the NMR
signal of the density operator (14) for one spin of the spin system at least
in the ensemble has a zero intensity. If the NMR signal of the density
operator $\rho _{f}$ of Eq.(14) for all spins in the ensemble have nonzero
intensities the above process can not determine certainly the parity
problem. To solve this problem one needs to modify the above quantum
algorithm. First, a phase-shift operation $U_{M}(-\pi /2)$ is introduced.
This known phase-shift unitary operation is not an oracle unitary operation
and is applied only on the work qubits $I$ of the ensemble. It is defined by

$\qquad \qquad \qquad U_{M}(-\pi /2)=\stackrel{M}{\stackunder{l}{\prod }}%
C_{l}(-\pi /2).\qquad \qquad \qquad \qquad \qquad \ \ \qquad (16)$ \newline
This phase-shift unitary operation is the product of $M$ selective
phase-shift operations $C_{l}(-\pi /2)$, while these selective phase-shift
operations are given in advance and are required to meet some constraint
that will be stated below. It turns out easily that the phase-shift unitary
operator $U_{M}(-\pi /2)$ commutes with the oracle unitary operator $%
U_{o}(\theta )$: $[U_{M}(-\pi /2),U_{o}(\theta )]=0.$ Now a modified quantum
algorithm to solve the parity problem is given below. One first makes an
oracle unitary operation of $U_{o}(\pi /2)$ on the initial density operator $%
\rho _{I}(0)$ of Eq.(11) and then applies the known phase-shift unitary
operation $U_{M}(-\pi /2)$ to the ensemble, then adds y-axis direction 90
degree pulse $90_{y}^{\circ }$ and a purge pulse just as the above process.
It is easy to prove along the line of the above process that the final
density operator now is given by

$\qquad \qquad \rho _{f}=(2/N)\stackrel{n}{\stackunder{k=1}{\sum }}%
\varepsilon _{k}(P_{k}-M_{k})I_{kz}\qquad \qquad \qquad \qquad \qquad \qquad
\ (17)$ \newline
where the contribution of the $M_{k}$ part to the NMR signal of Eq.(17)
comes from the phase-shift unitary operation $U_{M}(-\pi /2).$ The integer $%
M_{k}$ is given by

$\qquad \qquad \qquad \qquad M_{k}=\stackrel{M}{\stackunder{l}{\sum }}%
a_{k}^{l}.\qquad \qquad \qquad \qquad \qquad \quad \qquad \qquad \qquad (18)$
\newline
Note that unlike the integer $P_{k}$ each integer $M_{k}$ is given in
advance and can take any integer from $-N/2$ to $N/2$. Now the intensity of
the NMR signal of the spin $k$ of the spin system in the ensemble is
proportional to the subtraction $(P_{k}-M_{k})$ of the two integers $P_{k}$
and $M_{k}$ instead of the single integer $P_{k}$. The subtraction $%
(P_{k}-M_{k})$ can be positive, negative, or zero. Obviously, The
subtraction $(G-M)$ is certainly an even number if there is zero-intensity
NMR signal of the density operator $\rho _{f}$ of Eq.(17) for any arbitrary
spin $k$ of the spin system in the ensemble. In the case the parity of the
integer phase parameter $G$ can be determined directly from the subtraction $%
(G-M)$ since the integer $M$ is known , that is, $G$ is an even number if
the integer $M$ is an even number, otherwise $G$ is an odd number. How to
find the zero point of the subtraction $(P_{k}-M_{k})$ with the above
quantum algorithms? This is a classical one-dimensional search problem to
find a zero point in mathematics. At the first step the original quantum
algorithm is run once, that is, making the oracle unitary operation of $%
U_{o}(\pi /2)$ on the initial density operator of Eq.(11), then adding a
hard $90_{y}^{\circ }$ pulse on the work qubits $I$ and a purge pulse on the
ensemble, then applying a readout $90^{\circ }$ pulse on the work qubits $I$
and recording the NMR signal of work qubits $I$ during decoupling the
auxiliary qubits $S$ [14] by NMR phase-sensitive measurement [5, 6]. If one
can find that there is a zero-intensity NMR signal of any spin $k$ in the
fast Fourier transform NMR spectroscopy of the recorded NMR signal, it can
be certain that the integer phase parameter $G$ is an even number and the
parity function $par(f)=+1$, otherwise one needs to run further the above
modified quantum algorithm. Without losing generality, choose an arbitrary
spin $k$ of the spin system in the ensemble with $P_{k}>0$ (note that $%
N/2\geq P_{k}\geq -N/2$), then a suitable phase-shift operation $U_{M}(-\pi
/2)$ with $M_{k}=N/4$ is set up (if $P_{k}<0,$ $M_{k}=-N/4$)$.$ By running
the modified quantum algorithm one can find either $(P_{k}-M_{k})\neq 0$ or $%
(P_{k}+M_{k})=0$ from the fast Fourier transform NMR spectroscopy of the
recorded NMR signal of Eq.(17). If $(P_{k}-M_{k})=0$ the parity problem is
solved. If $(P_{k}-M_{k})>0$ a new phase-shift operation $U_{M}(-\pi /2)$
with $M_{k}=3N/8$ is constructed, and if $(P_{k}-M_{k})<0$ another
phase-shift operation $U_{M}(-\pi /2)$ with $M_{k}=N/8$ is built up. Then
running the modified quantum algorithm again one can solve the parity
problem if $(P_{k}-M_{k})=0,$ otherwise construct again a new phase-shift
unitary operation $U_{M}(-\pi /2)$, and then run the modified quantum
algorithm again till one finds the zero point $(P_{k}-M_{k})=0.$ Obviously,
the above search process really uses the classical one-dimensional two
partition method to search for a zero point. It needs $n$ times at most
repeat running the quantum algorithm to find the zero point $(P_{k}-M_{k})=0$
(note that $N=2^{n}$) by changing the number $M_{k}.$ Therefore, one can
solve efficiently the parity problem by repeating $n$ times at most to run
the above quantum algorithm. \newline
\newline
\textbf{5. The explicit construction of the phase-shift unitary operation }$%
U_{M}(-\pi /2)$\newline

The phase-shift operations $U_{M}(-\pi /2)$ is a diagonal unitary operator
and its effective Hamiltonian can be expanded in the longitudinal
magnetization and spin order operator subspace [22]. It is the product of $M$
selective phase-shift operations $C_{l}(-\pi /2),$ as shown in Eq.(16).
Without losing generality, one can set conveniently the index $k=1$ and $%
M=|M_{1}|$ in Eq.(18), then for all $M$ indexes $l$ in the sum of Eq.(18)
the unity numbers $\{a_{1}^{l}\}$ take the same value, i.e., $a_{1}^{l}=+1$ $%
(M_{1}>0),$ or $a_{1}^{l}$ $=-1$ $(M_{1}<0)$. Then the phase-shift unitary
operation $U_{M}(-\pi /2)$ in the unity-number vector representation $%
\{a_{k}^{s}\}$ may be expressed as, respectively,

$U_{M}(-\pi /2)=\exp [i\pi /2\stackrel{M}{\stackunder{l}{\sum }}(\frac{1}{2}%
E_{1}+I_{1z})\bigotimes (\frac{1}{2}E_{2}+a_{2}^{l}I_{2z})$

$\qquad \qquad \qquad \bigotimes ...\bigotimes (\frac{1}{2}%
E_{n}+a_{n}^{l}I_{nz})],\qquad \qquad \qquad \qquad \qquad \qquad \qquad
(19a)$ \newline
or

$U_{M}(-\pi /2)=\exp [i\pi /2\stackrel{M}{\stackunder{l}{\sum }}(\frac{1}{2}%
E_{1}-I_{1z})\bigotimes (\frac{1}{2}E_{2}+a_{2}^{l}I_{2z})$

$\qquad \qquad \qquad \bigotimes ...\bigotimes (\frac{1}{2}%
E_{n}+a_{n}^{l}I_{nz})].\qquad \qquad \qquad \qquad \qquad \qquad \quad \ \
(19b)$ \newline
Without losing generality only $U_{M}(-\pi /2)$ of Eq.(19a) with $%
a_{1}^{l}=+1$ for all index $l$ is constructed explicitly below. One simple
construction for the phase-shift unitary operations $U_{M}(-\pi /2)$ is
given explicitly by

$U_{1}(-\pi /2)=\exp [i\pi /2(\frac{1}{2}E_{1}+I_{1z})\bigotimes (\frac{1}{2}%
E_{2}+I_{2z})\bigotimes ...$

$\qquad \qquad \qquad \bigotimes (\frac{1}{2}E_{n-1}+I_{n-1z})\bigotimes (%
\frac{1}{2}E_{n}+I_{nz})],$

$U_{2}(-\pi /2)=\exp [i\pi /2(\frac{1}{2}E_{1}+I_{1z})\bigotimes (\frac{1}{2}%
E_{2}+I_{2z})\bigotimes ...\bigotimes (\frac{1}{2}E_{n-1}+I_{n-1z})],$

$U_{3}(-\pi /2)=\exp [i\pi /2(\frac{1}{2}E_{1}+I_{1z})\bigotimes (\frac{1}{2}%
E_{2}+I_{2z})\bigotimes ...\bigotimes (\frac{1}{2}E_{n-2}+I_{n-2z})]$

$\times \exp [i\pi /2(\frac{1}{2}E_{1}+I_{1z})\bigotimes (\frac{1}{2}%
E_{2}+I_{2z})\bigotimes ...\bigotimes (\frac{1}{2}E_{n-2}+I_{n-2z})]$

$\qquad \qquad \bigotimes (\frac{1}{2}E_{n-1}-I_{n-1z})\bigotimes (\frac{1}{2%
}E_{n}-I_{nz})],$

$.......$

$U_{2^{k}}(-\pi /2)=\exp [i\pi /2(\frac{1}{2}E_{1}+I_{1z})\bigotimes (\frac{1%
}{2}E_{2}+I_{2z})\bigotimes ...\bigotimes (\frac{1}{2}E_{n-k}+I_{n-kz})]$

......

$U_{N/2-1}(-\pi /2)=\exp [i\pi /2(\frac{1}{2}E_{1}+I_{1z})]$

$\qquad \times \exp [-i\pi /2(\frac{1}{2}E_{1}+I_{1z})\bigotimes (\frac{1}{2}%
E_{2}-I_{2z})\bigotimes ...\bigotimes (\frac{1}{2}E_{n}-I_{nz})],$

$U_{N/2}(-\pi /2)=\exp [i\pi /2(\frac{1}{2}E_{1}+I_{1z})]$\newline
where the unity operator $E_{k}$ is omitted without confusion in the direct
product terms, for example, $(\frac{1}{2}E_{1}+I_{1z})\bigotimes
...\bigotimes (\frac{1}{2}E_{n-1}+I_{n-1z})$ $=(\frac{1}{2}%
E_{1}+I_{1z})\bigotimes ...\bigotimes (\frac{1}{2}E_{n-1}+I_{n-1z})%
\bigotimes E_{n}.$ Obviously, besides the above construction there are a lot
of other different constructions for the phase-shift operations which can be
used to search for the zero point of $(P_{k}-M_{k})$ in the present quantum
algorithm. Because all these phase-shift unitary operations are not the
oracle unitary operation and are given in advance one can construct their
polynomial-time quantum circuits with the suitable interaction Hamiltonian
of the spin system. In all the phase-shift unitary operations $U_{M}(-\pi
/2) $ the basic unitary operations are $U_{2^{k}}(-\pi /2)=C_{0}^{n-k}(-\pi
/2)$ $(k=0,1,...,n-1)$, where $C_{0}^{n-k}(\theta )$ is a nonselective
phase-shift operation in the spin subsystem with $n-k$ qubits of the spin
system with $n$ qubits and is defined as $C_{0}^{n-k}(\theta )=\exp
(-i\theta D_{0}^{n-k})$ (note that $C_{0}^{n}(\theta )=C_{0}(\theta ))$ with
the diagonal operator $D_{0}^{n-k}=(\frac{1}{2}E_{1}+I_{1z})\bigotimes (%
\frac{1}{2}E_{2}+I_{2z})\bigotimes ...\bigotimes (\frac{1}{2}%
E_{n-k}+I_{n-kz}),$ $(k=0,1,...,n-1).$ It is well known that the
nonselective phase-shift operation $C_{0}^{n-k}(\theta )$ ($\theta =\pi $)
is the basic unitary operation in the Grover algorithm [15] with $2^{n-k}$
dimensional search space. It is easy to prove that any phase-shift operation 
$U_{M}(-\pi /2)$ $(1\leq M\leq N/2=2^{n-1})$ can be expressed as the product
of a polynomial number of the nonselective phase-shift operations $%
C_{0}^{n-k}(-\pi /2)$ $(k=0,1,...,n-1)$ and the single-qubit operations. As
an example, the phase-shift operation $U_{M}(-\pi /2)$ with $%
M=2^{k}+2^{l}+2^{0}$ $<2^{k+1}$ can be expressed explicitly as

$U_{M}(-\pi /2)=C_{0}^{n-k}(-\pi /2)\exp (-i\pi I_{n-kx})C_{0}^{n-l}(-\pi
/2)\exp (-i\pi I_{n-lx})$

$\qquad \qquad \qquad \times C_{0}(-\pi /2)\exp (i\pi I_{n-lx})\exp (i\pi
I_{n-kx})$\newline
In general case, the number $M$ can be expressed by the binary
representation as

$M=2^{k}+b_{k-1}2^{k-1}+b_{k-2}2^{k-2}+...+b_{1}2^{1}+b_{0}2^{0}<2^{k+1}$%
\newline
where $b_{l}=0,1;$ $l=0,1,...,k-1.$ Then the phase-shift operation $%
U_{M}(-\pi /2)$ can be generally written as\newline
$U_{M}(-\pi /2)=C_{0}^{n-k}(-\pi /2)\exp (-i\pi
I_{n-kx})C_{0}^{n-k+1}(-b_{k-1}\pi /2)\exp (-i\pi I_{n-k+1x})$\newline
$\times C_{0}^{n-k+2}(-b_{k-2}\pi /2)\exp (-i\pi
I_{n-k+2x})C_{0}^{n-k+3}(-b_{k-3}\pi /2)\exp (-i\pi I_{n-k+3x})...$

$\times C_{0}^{n-1}(-b_{1}\pi /2)\exp (-i\pi I_{n-1x})C_{0}(-b_{0}\pi
/2)\exp (i\pi I_{n-1x})...\exp (i\pi I_{n-kx})$\newline
Therefore, the phase-shift operation $U_{M}(-\pi /2)$ can be expressed as
the product of $(k+1)$ basic phase-shift operations $C_{0}^{n-k+l}(-\pi /2)$ 
$(l=0,1,...,k;$ $0\leq k\leq n-1)$ and $2k$ single-qubit unitary operations $%
\exp (\pm i\pi I_{n-k+lx})$ $(l=0,1,...,k-1)$ at most. \newline
\newline
\textbf{6. Discussion}\newline

It has showed in previous sections that the parity problem, one of the
hardest problems in quantum and classical computations, can be solved
efficiently on NMR quantum computers in polynomial time. It is based on the
quantum parallelism principle in a quantum ensemble, selective decoherence
manipulation, and the NMR phase-sensitive measurement. Because the parity
function is a collective quantity one need not find all the function values
so as to determine parity of the function on NMR quantum computers. In the
paper the collective quantity $G$ (see Eq.(8)) that reflects faithfully the
parity problem is contained in the oracle unitary operations $U_{f}$ and is
used to determine efficiently the parity of function. The main drawback of
the present quantum algorithm to solve the parity problem is that the
detectable NMR signal reduces exponentially as the qubit number. To
determine certainly the parity function with the present quantum algorithm
it is required that the detectable NMR signal-to-noise ratio of the spin
ensemble on an NMR machine must be as low as $2\varepsilon _{k}/N,$ as can
be seen in Eq.(14) or (17). These are similar to the basic characteristic
feature of the conventional NMR quantum computers based on the effective
pure states [9]. To implement really the present quantum algorithm one may
still pay attention to the NMR dynamic range problem on an NMR machine [5,
6]. This problem may be solved by using the spin ensemble of heteronuclear
spin system. The present quantum algorithm is polynomial-time one, while
those quantum algorithms running on the conventional quantum computers built
up with the quantum system [1, 2] or based on the effective pure states [7,
8] cannot solve efficiently the parity problem and cannot outperform the
classical algorithms. As an example, suppose that an NMR quantum computer
with $30$ qubits is available now [27]. Then to solve the parity problem
with 30 qubits the present quantum algorithm needs $\sim 30$ calls of the
oracle unitary operation $U_{o}(\pi /2)$ or about $60$ calls of the oracle
unitary operation $U_{f}.$ However, it will need $\sim 2^{30}/2$ $(\sim
5\times 10^{8})$ calls of the oracle unitary operation $U_{f}$ on the
conventional quantum computer and $2^{30}$ $(\sim 10^{9})$ calls of the
oracle unitary operation $U_{f}$ on a classical computer, respectively [1].
The difference of the call number of the oracle unitary operation is huge
between the present quantum algorithm and the conventional or the classical
one.

In comparison with the former scheme [14] to solve efficiently the parity
problem based on the spectral labelling method [23] the present quantum
algorithm is better one and more useful in practice. This is because the
present quantum algorithm can be run on the spin ensemble of the linear
molecules and does not required that the auxiliary qubits $S$ are coupled
with all the spins of the work qubits $I$, and moreover, it can use the
highly mix state of the spin ensemble instead of the effective pure state as
the initial input state.

One could enhance the NMR signal of the final density operators of Eqs.(14)
and (17) by running a quantum algorithm consisting of a polynomial number of
the oracle unitary operations and nonselective unitary operations instead of
the present quantum algorithm. However, the decoherence effect in a spin
ensemble limits the possible call number of the oracle unitary operation in
the quantum algorithm. Thus, there is a maximum call number of the oracle
unitary operation in the quantum algorithm or a maximum run time $T_{\max }$
of the quantum algorithm in the spin ensemble due to the decoherence effect.
The NMR signal will decrease simply but not be enhanced again as increasing
the call number of the oracle unitary operation in the quantum algorithm if
the call number is larger than the maximum one. This is because at that case
running the quantum algorithm will take more time than $T_{\max }$, so that
the decoherence effect becomes the dominant effect negatively upon the NMR
signal. If the NMR signal is still too weak to be detected on an NMR machine
when the quantum algorithm is run with the maximum number calls of the
oracle unitary operation, then other NMR signal enhancement methods such as
the polarization transfer techniques, etc., [5, 6, 27-29] could be used to
enhance the NMR signal and reach the same goal. Thus, here such NMR signal
enhancement methods could be an effective scheme to overcome the decoherence
effect on the quantum computation. \newline
\newline
\textbf{References}\newline
1. E.Farhi, J.Goldstone, S.Gutmann, and M.Sipser, Phys.Rev.Lett. 81, 5442
(1998) \newline
2. R.Beals, H.Buhrman, R.Cleve, M.Mosca, and R. de Wolf,

http://xxx.lanl.gov/abs/quant-ph/9802049 (1998)\newline
3. H.Buhrman, R.Cleve, and A.Wigderson,
http://xxx.lanl.gov/abs/quant-ph/9802040 (1998)\newline
4. Y.Ozhigov, http://xxx.lanl.gov/abs/quant-ph/9712051 (1997)\newline
5. R.R.Ernst, G.Bodenhausen, and A.Wokaun, Principles of nuclear magnetic
resonance in one and two dimensions (Oxford University Press, Oxford, 1987) 
\newline
6. R.Freeman, Spin Choreography, Spektrum, Oxford, 1997\newline
7. D.G.Cory, A.F.Fahmy, and T.F.Havel, Proc.Natl.Acad.Sci. USA 94, 1634
(1997)\newline
8. N.A.Gershenfeld and I.L.Chuang, Science 275, 350 (1997) \newline
9. W.S.Warren, Science 277, 1688 (1997) \newline
10. J.A.Jones, Fort.der.Physik. 48, 909 (2000)\newline
11. S.L.Braunstein, C.M.Caves, R.Jozsa, N.Linden, S.Popescu, and R.Schack,
Phys.Rev.Lett. 83, 1054 (1999) \newline
12. N.Linden and S.Popescu, http://xxx.lanl.gov/abs/quant-ph/9906008 (1999) 
\newline
13. R.Schack and C.M.Caves, Phys.Rev. A 60, 4354 (1999)\newline
14. X.Miao, http://xxx.lanl.gov/abs/quant-ph/0101126 (2001) \newline
15. L.K.Grover, Phys.Rev.Lett. 79, 325 (1997) \newline
16. I.L.Chuang, N.Gershenfeld, and M.Kubinec, Phys.Rev.Lett. 80, 3408 (1998)%
\newline
17. D.Deutsch, Proc.Roy.Soc.Lond. A 400, 97 (1985) \newline
18. C.H.Bennett, E.Bernstein, G.Brassard, and U.Vazirani, SIAM Journal on
Computing 26, 1510 (1997) \newline
19. H.Bennett, IBM J.Res.Develop. 17, 525 (1973) \newline
20. P.Benioff, J.Stat.Phys. 22, 563 (1980) \newline
21. R.Cleve, A.Ekert, C.Macchiavello, and M.Mosca, Proc.Roy.Soc.Lond. A 454,
339 (1998)\newline
22. (a) X.Miao, Mol.Phys. 98, 625 (2000)

(b) X.Miao, http://xxx.lanl.gov/abs/quant-ph/003068 (2000)\newline
23. Z.L.Madi, R.Bruschweiler, and R.R.Ernst, J.Chem.Phys. 109, 10603 (1998) 
\newline
24. A.L.Davis, G.Estcourt, J.Keeler, E.D.Laue, and J.J.Titman, \newline
J.Magn.Reson. A105, 167 (1993)\newline
25. O.W.Sorensen, M.Rance, and R.R.Ernst, J.Magn.Reson. 56, 527 (1984) 
\newline
26. X.Miao http://xxx.lanl.gov/abs/quant-ph/008094 (2000)\newline
27. N.Gershenfeld and I.L.Chuang, Science 277, 1689 (1997) \newline
28. L.J.Schulman and U.V.Vazirani, in Proceedings of the 31st Annual ACM
Symposium on Theory of Computing, 1999, pp.322\newline
29. P.O.Boykin, T.Mor, V.Roychowdhury, F.Vatan, and R.Vrijen,

http://xxx.lanl.gov/abs/quant-ph/0106093\newline

\end{document}